\documentclass{ws-mpla}
\usepackage[super]{cite}
\usepackage{graphicx}
\usepackage{epstopdf}
\usepackage{floatrow}
\usepackage{float}
\usepackage{amsmath,amssymb}
\usepackage{xcolor}
\usepackage{braket}
\usepackage{esvect}
\usepackage{tensor}
\usepackage{ulem}
\usepackage[theorems,breakable]{tcolorbox}
\allowdisplaybreaks

\begin{document}

\newcommand{\p}{\partial}
\newcommand{\hh}{{\widehat{h}}}
\newcommand{\bchi}{{\bar{\chi}}}
\newcommand{\btheta}{{\bar{\theta}}}
\newcommand{\ds}{{\slashed{\nabla}}}
\newcommand{\unit}{{\mathbbm{1}}}
\newcommand{\Tr}{{\rm Tr}}

\newcommand{\gb}{\bar{g}}
\newcommand{\Db}{\bar{D}}
\newcommand{\Rb}{\bar{R}}

\newcommand{\cO}{\mathcal{O}}
\newcommand{\cR}{\mathcal{R}}

\newcommand{\be}{\begin{equation}}
\newcommand{\ee}{\end{equation}}

\newcommand{\colF}[1]{\bgroup\color{blue}#1\egroup}
\newcommand{\coled}[1]{\textcolor{red}{\sout{#1}}}

\catchline{}{}{}{}{}

\title{A Information-Theoretic View on Spacetime}

\author{\footnotesize Frank Saueressig}

\address{Institute for Mathematics, Astrophysics and Particle Physics (IMAPP) \\ Radboud University, Heyendaalseweg 135 \\ Nijmegen, 6525 AJ,The Netherlands\footnote{f.saueressig@science.ru.nl}\\
author@emailaddress}

\author{Amir Khosravi}

\address{Institute for Mathematics, Astrophysics and Particle Physics (IMAPP) \\ Radboud University, Heyendaalseweg 135\\
Nijmegen, 6525 AJ,The Netherlands
}

\maketitle


\begin{abstract}
We argue in a quantitative way that the unitarity principle of quantum field theory together with the quantum information bound on correlation functions are in tension with a space which is made out of disconnected patches at microscopic scales.

\end{abstract}

\ccode{PACS Nos.: 04.60.-m, 11.10.Hi, 03.65.Ud}

\section{Introduction}	

The quest to unravel the nature of space and time is one of the oldest and most challenging ones in the history of intellectual endeavors. Quite intriguingly, combining fundamental principles of quantum mechanics with an information theoretic approach to spacetime allows making statements about the structure of space and time way beyond experimentally accessible scales \cite{spin,string,string2}. In this work we use arguments based on entanglement entropy and unitarity to conclude that spatial slices of spacetime must be path-connected topological spaces in order to avoid running into conflicts with these concepts.

In Sect.\ \ref{II.1} we review the notion of entanglement entropy and quantum mutual information. In Sect.\ \ref{II.2} we derive a constraint that is set by the unitarity principle of quantum mechanics on the mutual information shared by two regions of space. In Sect.\ \ref{II.3}, we argue that space cannot consist of (microscopic) disconnected structures without violating this principle. Finally, a summary is provided.

\section{Quantum Information, Entanglement and Spacetime} \label{II}

\subsection{A Brief Introduction} \label{II.1}

Consider a 3-dimensional spatial region $C$ together with its complement $D$ in a 4-dimensional spacetime. Thus $C$ and $D$ are disjoint except for the 2-dimensional boundary shared by them. Suppose the total information about the geometry of region $E=C\cup D$ is stored in a density matrix $\rho_{\rm CD}$. The entangelement entropy between the two regions can then be defined via the von Neumann formula as
\begin{equation}\label{ent}
S_{C} =-  \Tr [\rho_{C} \log{\rho_{C}}] \, \, , \qquad  S_{D} =-  \Tr [\rho_{D} \log{\rho_{D}}] \, . 
\end{equation}
Here $\rho_{C}$ and $\rho_{D}$ are the reduced density matrices constructed out of $\rho_{CD}$ by tracing out $D$ and $C$:
\begin{equation}\label{rho}
\rho_{C} =  \Tr_{D} [\rho_{CD}] \, \, , \qquad  \rho_{D} =  \Tr_{C} [\rho_{CD}] \, . 
\end{equation}
The mutual information $I_{\rm CD}$ is given by \cite{qminfo}
\begin{equation}\label{main}
I_{\rm CD}\equiv S_{C}+S_{D}-S_{\rm CD} \geq 0
\end{equation}
where $S_{\rm CD} \equiv -  \Tr [\rho_{CD} \log{\rho_{CD}}]$ is the total entanglement of the $(C ,\, D)$-system. In a general state, $I_{CD}$ is a measure of correlation between $C$ and $D$. It essentially indicates how much region $C$ knows about $D$ and vice versa. If all the correlation is quantum, i.e. $S_{\rm CD}=0$, then we have a pure state $S_{C}=S_{D}$.

\subsection{Unitarity and Mutual Information} \label{II.2}

The question we seek to answer is: "How does unitarity constrain the mutual information shared by two regions of space?". To address this question we start from the Callan-Symanzik equation \footnote{For example one could consider a massive scalar field living on a predescribed background spacetime.}. This equation implies the following differential equation for $n$-point correlation functions $\Gamma^{(n)}(p_j, g(\mu), \mu)$ \cite{mag}
\begin{equation} \label{cz}
\left[\mu\,\partial_{\mu}+\beta(g(\mu)) \frac{\partial}{\partial {g(\mu)}}-\frac{n}{2}\eta\left(g(\mu)\right) \right]\Gamma^{(n)}(p_j, g(\mu), \mu)=0 .
\end{equation}
Here $\mu$ is the renormalization scale, $\eta\left(g(\mu)\right)$ is the anomalous dimension of the fundamental fields appearing in the action, $g(\mu)$ is a set of coupling constants and $g(\mu)$ satisfies the renormalization group equation $\beta(g(\mu))=\mu \frac{d}{d \mu} g(\mu)$. The background field formalism allows to extend these renormalization group techniques also to the deep quantum gravity regime \cite{Reuter:1996cp} by treating quantum fluctuations of the spacetime metric akin to a quantum field theory in curved spacetime \cite{Niedermaier:2006wt,robertobook}. The general solution to ``\eqref{cz}'' can be written as 
\begin{equation} \label{solcz}
\Gamma^{(n)}(p_j, g(\mu), \mu)=exp\left[\frac{n}{2}\int_{\mu_0}^{\mu}\frac{d\mu^{'}}{\mu^{'}}\eta\left(g(\mu^{'})\right)\right]\, \Gamma^{(n)}(p_j, g(\mu_0), \mu_0) \, . 
\end{equation}

The crucial observation is that for a generic \textbf{unitary} quantum field theory (QFT) it is safe to assume $\eta(g(\mu))\geq 0$ \cite{unit,unit2}. If we then define operators $O_C$ and $O_D$ with support on $C$ and $D$ respectively, the connected two-point function $\langle O_C O_D \rangle_c$ is positive semi-definite $\braket{O_{C}\,O_{D}}_c  \geq 0$. Consequently, using ``\eqref{solcz}'', we get $\partial_{\mu}\braket{O_{C}\,O_{D}}_c \geq 0$. A rigorously proven example of this feature is the celebrated c-theorem in conformal field theory where unitarity implies the positivity of the central charge and consequently $\partial_{\mu}\braket{O_{C}\,O_{D}}_c \geq 0$ \cite{cft,cft2}.

On the other hand, for any two operators $O_{C}$ and $O_{D}$, one can use the well-established properties of the conditional probabilities and certain results of quantum information science to show that the mutual information between the two regions $C$ and $D$ obeys \cite{qminfo1}:
\begin{equation}\label{infobound}
I_{\rm CD} \geqslant \frac{\braket{O_{C}\,O_{D}}_c^2}{2|O_C|^2\,|O_{D}|^2}.
\end{equation}
In more physical terms, ``\eqref{infobound}'' indicates that the quantum mutual information shared by two regions of space is the maximum amount of correlation they can have with each other. 

Regarding ``\eqref{infobound}'' it is crucial to distinguish two  cases: firstly, ``\eqref{infobound}'' can be applied to "discrete" spaces coming with a hard UV-cutoff. Such spaces would be path-disconnected by definition and the operator norms $|O|$ are finite by construction. This is the case considered in van Raamsdonk's work \cite{van}, where it was concluded that if the mutual information between $C$ and $D$ goes to zero, then all correlations must decrease to zero also. Secondly, one can consider continuum space which is path connected and leads to infinite values of $|O|$. As a consequence vanishing mutual information $I_{CD}$ does not automatically imply the vanishing of the correlation function. In the sequel,  we will work in the first setting and construct a contradiction between the unitarity principle and a discrete space.

Any physically interesting theory should have at least one correlation function which has support on both $C$ and $D$ and is non-zero at some (finite) scale. Focusing on such a correlation function, we can make the following argument: assume the value of shared information between regions $C$ and $D$ is zero in the ultraviolet (UV) limit, i.e. we have $I_{\rm CD}^{UV} = 0$. From unitarity we have $\braket{O_{C}\,O_{D}}_c \geq 0$ and thus eq.\ ``\eqref{infobound}'' entails that $\langle O_C O_D \rangle_c^{UV} = 0$. Furthermore, if the unitarity principle is satisfied, we should have $\partial_{\mu}\braket{O_{C}\,O_{D}}_c \geq 0$. If we take $\partial_{\mu}\braket{O_{C}\,O_{D}}_c = 0$ then the correlation function vanishes on all scales contradicting our premise above. So unitarity and the existence of non-vanishing correlations imply $\partial_{\mu}\braket{O_{C}\,O_{D}}_c > 0$. This results in $\langle O_C O_D \rangle_c^{\rm IR} < 0$ which conflicts with unitarity. The bottom line of this argument is that the requirement of unitarity and eq.\ ``\eqref{infobound}'' implies $I_{\rm CD}^{UV} > 0$. In Sect.\ \ref{II.3} we argue that $I_{\rm CD}^{UV} > 0$ is in conflict with a disconnected space.\footnote{ Our statement is actually compatible with the situation where both regions $C$ and $D$ support non-trivial, unitary QFTs which do not interact with each other. In this case the two-point correlator $\langle \cO_C \, \cO_D \rangle = \langle \cO_C \rangle \langle \cO_D \rangle$ factorizes, implying the vanishing of the connected correlation function $\langle \cO_C \cO_D \rangle_c = \langle \cO_C \, \cO_D \rangle  - \langle \cO_C \rangle \langle \cO_D \rangle \stackrel{!}{=} 0$.}

An alternative derivation of this conclusion builds on the $I_{theorem}$ \cite{I-theorem}. In this case one can use $\partial_{\mu} I_{CD} > 0$ to argue that $I_{CD}^{UV} = 0$ implies $I_{CD}^{IR} < 0$ and hence leads to the violation of entanglement subadditivity \footnote{Subadditivity of the entanglement of two regions $C$ and $D$ means $\rm{Entanglement}_{C \cup D} \leq \rm{Entanglement}_{C}+\rm{Entanglement}_{D}$. The entanglement entropy always satisfies this property \cite{qminfo}.} in the IR. At this stage, there is no consensus on the existence of the $I_{theorem}$ though.

\subsection{Space Must be Path-Connected} \label{II.3}

According to Van Raamsdonk \cite{van} the relation between the entanglement entropy of two regions of space, their shared area and the geodesic distance between them has a very remarkable interpretation. The original argument is presented in an AdS/CFT \cite{adscft,van} setup; here we present a simpler version using properties of the entanglement entropy in QFT. References \cite{sred,sred2,bia} suggest that to leading order in the UV cutoff the entanglement entropy is proportional to  
the shared area (area of the entangling surface) between a region and its complement in space. Thus, reducing the entanglement entropy and mutual information is equivalent to reducing the shared area. 

Furthermore, for a typical massive field theory with mass $m$ we have $\braket{O_{C}(\vec{x}) \,O_{D}(\vec{y})}_c \propto e^{-m\, L}$ where $\vec{x}$ and $\vec{y}$ are spatial positions in $C$ and $D$ and $L$ indicates the geodesic distance between these points. In the light of ``\eqref{infobound}'', this means $I \ll 1$ if $L \gg \frac{1}{m}$. 

Combining last two paragraphs, one can summarize Van Raamsdonk's result as \cite{van}\footnote{Notably, the holographic entanglement entropy upon which the original Van Raamsdonk argument is based produces the ordinary QFT area law to leading order in the UV cutoff \cite{holo}.}:
\begin{equation}\tag{S1}\label{S1}
\parbox[c]{0.8\linewidth}{\it Zero entanglement entropy is equivalent to $C$ and $D$ being disconnected regions of space. Hence, the connectivity of the classical geometry has its roots in the properties of quantum entanglement entropy.}
\end{equation}

The core of our argument, essentially Sect.\ \ref{II.2}, survives the transition to a fundamental theory of nature as long as this theory admits an effective QFT description that is unitary and respects the basic rules of quantum information physics. If QFT itself survives the generalization, we have a scenario like asymptotic safety \cite{asym,robertobook,martin2}. If QFT is an effective description, we have something like a string theory effective description of the bulk \cite{adscft,string}. Regarding a disconnected space, according to ``\eqref{S1}'' we should have $I_{CD}^{UV}=0$ if regions $C$ and $D$ are disconnected atoms of space sharing no boundary. On the other hand, we know from the previous section that according to ``\eqref{infobound}'' the only way to reconcile $I_{CD}^{UV}=0$ with unitarity is by all correlation functions involving products of operators with support on $C$ and $D$ vanishing at all scales. If the disconnected regions are microscopic, this is of no interest in any physics scenario. So if one insists on a disconnected space in the UV and a QFT describing the universe, one has only two options: 1) Accept unitarity and consequently violate ``\eqref{infobound}'' which means certain basic rules of quantum information physics are violated. 2) Drop the unitarity principle and accept the possibility of information loss. Both options are in tension with principles of quantum mechanics. This suggests that space cannot be disconnected. A prototypical example for a physical process ruled out by these considerations is the creation and pinching-off of baby universes from our spacetime: if a quantum field has support on the region of spacetime that is pinched-off the information is lost entailing a violation of unitarity.

\section{Summary} \label{III}

In this manuscript we assumed that the universe admits a unitary QFT description, at least at the effective level. On this basis we argued that such a description together with rules of conditional probability theory and quantum information science implies that space cannot emerge from the coarse graining of a patchwork of disconnected structures. This is a consequence of ``\eqref{S1}'', the unitarity principle, ``\eqref{infobound}'' and the requirement of a non-trivial correlation function. To put it in other words, in the light of ``\eqref{S1}'' it is obvious that $I_{CD}^{UV} = 0$ is a necessary condition for having a disconnected space. Thus unitarity and ``\eqref{infobound}'' put conditions on the topological properties of space, demanding $I_{CD}^{UV} > 0$. Notably, our analysis does not imply that on a space made of path-disconnected regions, any unitary QFT is necessary trivial: every path-connected region may support its own unitary QFT coming with non-trivial correlation functions. If the regions are Planck-sized this is not a physically interesting scenario though.

An interesting situation motivating a non-trivial generalization of the ideas advocated in this work is the BKL scenario \cite{BKL}. In this model spatial points decouple close to the cosmic singularity. This should lead to vanishing correlation functions in the UV. At the same time, spacetime remains connected, thereby violating the assumption of fundamental discreteness made in our derivation. It would be interesting to understand the interplay between the information theoretic concepts underlying our work and unitarity in such a situation as well. We also note that our argument may not be applicable to (apparently discrete) models of quantum gravity as, e.g., causal dynamical triangulations\cite{Ambjorn:2012jv,Loll:2019rdj} 
or group field theories \cite{Freidel:2005qe,Carrozza:2016vsq},
since in this case the ``discreteness scale'' plays the role of a UV regulator which is to be removed by taking a suitable continuum limit.

As a final remark, we note that the incompatibility between the unitarity of QFT and ``\eqref{infobound}'' with a disconnected space only depends on a set of basic properties of conditional probabilities and the unitarity of QFT. This result may restrict the class of (quantum) theories of gravity that are compatible with unitarity and basic rules of quantum information theory. 

\section*{Acknowledgement}

We would like to thank Carlo Rovelli, Slava Rychkov, Tatsuma Nishioka and Horacio Casini for fruitful discussions and Francesco Caravelli, Andrea Stergiou and Andrea Trombettoni for critical comments. Special thanks to Alfio Bonanno for providing insightful comments on the draft. A.K.\ profited from illuminating discussion during the {\it Functional and Renormalization-Group methods} \text{(FRGIM 2019)} conference in Trento, Italy where the work was presented. Attending the FRGIM 2019 was made possible by the generous financial support from the “ACRI-Young Investigators Training Program”.


\begin{thebibliography}{99}
	
\bibitem{spin} C.\ Rovelli, L.\ Smolin, “Spin Networks and Quantum Gravity”, Phys.\ Rev.\ D52 (1995) 5743, gr-qc/9505006.

\bibitem{string} 
G.\ T.\ Horowitz, “Spacetime in string theory”, New J.\ Phys.\ 7 (2005) 201, gr-qc/0410049.

\bibitem{string2}
N.\ Seiberg, “Emergent Spacetime”, Rapporteur talk at the 23rd Solvay Conference in Physics (2005), hep-th/0601234.	

\bibitem{qminfo} M.\ A.\ Nielsen, and I.\ L.\ Chuang, “Quantum computation and quantum information”, Camb.\ Univ.\ Press (Cambridge, UK) 2000.

\bibitem{mag} M.\ Maggiore, “A Modern Introduction to Quantum Field Theory”, Oxford University Press (Oxford, UK) 2005.

\bibitem{Reuter:1996cp}
M.~Reuter,
``Nonperturbative evolution equation for quantum gravity,''
Phys. Rev. D \textbf{57} (1998), 971,
arXiv:hep-th/9605030.

\bibitem{Niedermaier:2006wt}
M.~Niedermaier and M.~Reuter,
``The Asymptotic Safety Scenario in Quantum Gravity,
Living Rev. Rel. \textbf{9} (2006) 5.

\bibitem{robertobook}  R.\ Percacci, “An Introduction to Covariant Quantum Gravity and Asymptotic Safety”, World Scientific (Singapore) 2017.

\bibitem{unit} K.\ Higashijima, E.\ Itou, “Unitarity Bound of the Wave Function Renormalization Constant”, Prog.\ Theor.\ Phys.\ 110 (2003) 107, arXiv:0304047

\bibitem{unit2}
O.\ J.\ Rosten, “Triviality from the Exact Renormalization Group”, JHEP 07 (2009) 019, arXiv:0808.0082.

\bibitem{cft} A.\ B.\ Zamolodchikov, “Irreversibility of the Flux of the Renormalization Group in a 2D Field Theory”, JETP Lett.\ 43 (1986) 730.

\bibitem{cft2}
J.\ Cardy, "Is there a c-theorem in four dimensions?", Phys.\ Lett.\ B215 (1988) 749. 

\bibitem{qminfo1} M.\ M.\ Wolf, F.\ Verstraete, M.\ B.\ Hastings, J.\ I.\ Cirac, “Area laws in quantum systems: mutual information and correlations,” Phys.\ Rev.\ Lett.\ 100 (2008) 070502, arXiv:0704.3906.

\bibitem{adscft} J.\ M.\ Maldacena, “The Large N Limit of Superconformal Field Theories and Supergravity”, Adv.\ Theor.\ Math.\ Phys.\ 2 (1998) 231 [Int.\ J.\ Theor.\ Phys.\ 38 (1999) 1113], arXiv:hep-th/9711200.

\bibitem{van} M.\ Van Raamsdonk, “Building up spacetime with quantum entanglement”, Gen.\ Rel.\ Grav.\ 42 (2010) 2323, Int.\ J.\ Mod.\ Phys.\ D19 (2010) 2429, arXiv:1005.3035.

\bibitem{I-theorem} Y.\ Nakaguchi, T.\ J.\ Nishioka, “Entanglement Entropy of Annulus in Three Dimensions”, JHEP 04 (2015) 072, arXiv:1501.01293.

\bibitem{sred} M. Srednicki, “Entropy and Area”, Phys.\ Rev.\ Lett.\ 71 (1993) 666, arXiv:9303048.

\bibitem{sred2}
L.\ Bombelli, R.\ K.\ Koul, J.\ Lee, R.\ D.\ Sorkin, “Quantum source of entropy for black holes”, Phys.\ Rev.\ D34 (1986) 373.

\bibitem{bia} E.\ Bianchi, R.\ C.\ Myers, “On the Architecture of Spacetime Geometry”,     Class.\ Quant.\ Grav.\ 31 (2014) 214002, arXiv:1212.5183.

\bibitem{holo} T.\ Nishioka, S.\ Ryu and T.\ Takayanagi, “Holographic Entanglement Entropy: An Overview”, J.\ Phys.\ A 42 (2009) 504008, arXiv:0905.0932.

\bibitem{asym} S. Weinberg, “Critical Phenomena for Field Theorists”, In: A. Zichichi (eds) “Understanding the Fundamental Constituents of Matter”, The Subnuclear Series, vol 14. Springer (Boston, MA) 1978.

\bibitem{martin2}
M.\ Reuter and F.\ Saueressig, “Quantum Gravity and the Functional Renormalization Group”, Camb.\ Univ.\ Press (Cambridge, UK) 2019.

\bibitem{BKL}  V.\ A.\ Belinskii, I.\ M.\ Khalatnikov, E.\ M.\ Lifshitz, “Oscillatory approach to a singular point in the relativistic cosmology”, Adv.\ Phys.\ 19 (1970) 525.

\bibitem{Ambjorn:2012jv}
J.~Ambjorn, A.~Goerlich, J.~Jurkiewicz and R.~Loll,
``Nonperturbative Quantum Gravity,''
Phys. Rept. \textbf{519} (2012) 127,
arXiv:1203.3591 [hep-th].

\bibitem{Loll:2019rdj}
R.~Loll,
``Quantum Gravity from Causal Dynamical Triangulations: A Review,''
Class. Quant. Grav. \textbf{37} (2020) 013002,
arXiv:1905.08669 [hep-th].

%
\bibitem{Freidel:2005qe}
L.~Freidel,
``Group field theory: An Overview,
Int. J. Theor. Phys. \textbf{44} (2005) 1769,
arXiv:hep-th/0505016.

\bibitem{Carrozza:2016vsq}
S.~Carrozza,
``Flowing in Group Field Theory Space: a Review,''
SIGMA \textbf{12} (2016) 070,
arXiv:1603.01902 [gr-qc].


\end{thebibliography}
\end{document}